# "Seeing Sound": Audio Classification using the Wigner-Wille Distribution and Convolutional Neural Networks

Christonasis Antonios Marios[1], Stef van Eijndhoven[2], and Peter Duin[3]

[1] Engineering Doctorate Data Science, Technical University of Eindhoven (TU/e), Eindhoven, The Netherlands
[2] Engineering Doctorate Data Science, Technical University of Eindhoven (TU/e), Eindhoven, The Netherlands
[3] Dutch National Police, Eindhoven, The Netherlands

E-mail(s): [1] antwnchris@gmail.com
LinkedIn: [1] https://www.linkedin.com/in/amchristonasis/

**Abstract**

With big data becoming increasingly available, IoT hardware becoming widely adopted, and AI capabilities becoming more powerful, organizations are continuously investing in sensing. Data coming from sensor networks are currently combined with sensor fusion and AI algorithms to drive innovation in fields such as self-driving cars. Data from these sensors can be utilized in numerous use cases, including alerts in safety systems of urban settings, for events such as gun shots and explosions. Moreover, diverse types of sensors, such as sound sensors, can be utilized in low-light conditions or at locations where a camera is not available. This paper investigates the potential of the utilization of sound-sensor data in an urban context.

Technically, we propose a novel approach of classifying sound data using the Wigner-Ville distribution and Convolutional Neural Networks. In this paper, we report on the performance of the approach on open-source datasets. The concept and work presented is based on my doctoral thesis, which was performed as part of the Engineering Doctorate program in Data Science at the University of Eindhoven, in collaboration with the Dutch National Police. Additional work on real-world datasets was performed during the thesis, which are not presented here due to confidentiality.

## 1. Introduction

In the past years, Deep Learning has made enormous steps in Image Recognition. Open-source image datasets are abundant and architectures like Convolutional Neural Networks have been used to achieve performance that can even surpass human accuracy [1].

However, the same is not the case for sound-sensor data. While sensor networks, IoT hardware, and 5G are becoming increasingly part of our urban settings, research on the application of AI and sensor fusion on such settings has been more scarce. Open-source datasets are much more limited and techniques to classify this data are still in research phase.

Datasets like AudioSet [2] and techniques using the log-mel spectrogram and Convolutional Neural Networks [3] are currently the state of the art. That being said, there is still a lot of untapped potential, both in terms of research, as well as real-world applications.

In this paper, we provide a collection of open-source sound datasets that we encountered in our research, some of which proved useful in assessing the performance of our approach. Moreover, we propose a novel approach to classify sound clips, using the Wigner-Ville time-frequency distribution and Convolutional Neural Networks, that we have not seen anywhere else in the literature. The advantage of the Wigner-Ville distribution compared to log-mel spectrograms are the more detailed resolution it can provide in the time-frequency spectrum and a more accurate estimate of the instantaneous frequency [4].

## 2. Open-Source Datasets

A vital and challenging step of this research was the identification and collection of relevant open-source datasets.

It was interesting to observe that open-source datasets in this area of research were limited, especially when compared to many available open datasets containing structured image data. Despite this lack of public datasets containing acoustical sound signals, we were able to identify open-source datasets that could be used for training sound classifiers. The most notable are presented as follows.

### 2.1 AudioSet

AudioSet [2] is open sourced by Google Research. It contains around two million audio clips with a duration of ten seconds; they are fetched from YouTube videos and labelled semi-automatically. The dataset is weakly labeled, meaning that we only know the presence of a sound class in each sound clip, but we do not know exactly when during the ten seconds the sound of that class is apparent. Finally, the dataset is suited for a multi-label classifier, since more than one sounds can be present in each sound clip, whereas for this research we aimed to design a multi-class classifier. [5]

From the two million sound clips, two structured subsets were made available by Google Research for training and validation. The training set contains around 20,000 clips and the validation set around 15,000 clips. The limitation with AudioSet is that Google Research provides only the 128-dimensional feature vector they computed for each clip based on log-mel spectrograms [6]. For that reason, someone must scrape the raw audio clips from YouTube, if they want to apply a different transformation to the sound signal, as was the case in this research. That is a very time-consuming process and, even more, several YouTube videos from the original datasets have been removed by YouTube. As a result, there is a small reduction in the number of audio clips someone can scrape, compared to the original datasets.

### 2.2 UrbanSound8K

The UrbanSound8K dataset [7] is open sourced by researchers from the New York University. The clips were scraped from the freesound.org website and were labelled manually. The dataset contains around 8000 clips with a duration ranging from 4 to 30 seconds. The clips belong to 10 classes corresponding to common urban sounds. The dataset is provided in the form of raw audio clips. In addition, each audio clip contains one distinct sound, which makes it suitable for training multi-class classifiers.

### 2.3 Urban-SED

The Urban-SED dataset [8] is suitable for building sound segmentation algorithms. It contains ten thousand ten-second clips which are created as follows. Background noise combined with smaller audio clips from the UrbanSound8K dataset are present during each ten-second clip. Annotations for the sound classes and their start and end times are available.

### 2.4 Environmental Sound Classification-50 (ESC-50) Dataset

The ESC-50 dataset [9] contains four-second clips from fifty different classes of environmental sounds, ranging from animal, to human, to machine sounds. Each class features forty available clips. In total, the dataset contains 20000 audio clips. This dataset has been used for benchmarking many different sound classification algorithms. It is useful for creating multi-class classifiers in the context of environment understanding.

### 2.5 Montevideo Audio Visual Dataset (MAVD-Traffic)

MAVD-Traffic [10] is another dataset that is unique and was recently open-sourced. The dataset comes from the Montevideo University in Uruguay. The creators recorded both audio and video from different locations near roads in Montevideo city. They also provided annotations for the different vehicles that pass by with their start and end times. The annotations are provided at different levels based on a taxonomy they defined. In total, the audio and video recordings last around four hours.

The downside of the Montevideo recordings is that cars are passing by most of the time, whereas other vehicles pass by with a much lower frequency. The result is a very unbalanced dataset that is also quite noisy in terms of traffic noise.

Furthermore, multiple vehicles can pass by at the same time, which results in sounds getting mixed up. Finally, the annotations are weak, in the sense that they are not detailed as to exactly when each individual vehicle passes by the recorders. That being said, it is a unique dataset and very close to a real-life scenario.

*2.6 Military Vehicle Dataset*

Last but not least, the military vehicle dataset was collected in November 2001 in the city of Twenty-Nine Palms in California [11]. The authors collected data from sound, seismic, and infrared sensors, while two military vehicles were driving around an area of 900 x 300 m2, where a wireless sensor network of the abovementioned sensors was deployed. The military vehicles were a dragon wagon (209 entries) and an assault amphibian vehicle (180 entries). All the recordings ranged from 1.5 second to around 38 seconds with an average of around 13 seconds. The acoustic data was recorded at a rate of 4960 Hz. Each sound recording was annotated with the name of the vehicle that passed by the sensor node.

## 3. Literature Review

Approaches that are currently considered state-of-the-art follow the below-mentioned general classification strategy:

1. Transformation of the sound signals to images (arrays) with the use of time-frequency analysis techniques, such as the spectrogram and the mel-frequency cepstral coefficients (MFCCs).
2. Classification with the use of image classification techniques. The state of the art here is neural network architectures that make use of convolutional layers.

Several research papers have been published in the past few years in which approaches similar to the abovementioned strategy are proposed. For the case of the current research, the most influential papers were the AudioSet [2] and VGGish [3] papers from Google Research. In the AudioSet paper, the authors describe the creation of a generic dataset that can become an equivalent to ImageNet for sound research. More details about AudioSet can be found in the previous section. In the technical approach part, the VGGish paper was the most influential in our approach. In this paper, the authors propose a neural network architecture based on convolutional layers that is inspired by the popular VGG architecture [12] for image classification.

## 4. Time-Frequency Distributions and the Wigner-Ville Distribution

Time-Frequency Distributions are a very useful tool to perform analyses of signals. They provide a way to analyze signals both at the temporal and frequency domain at the same time. Essentially, they show the dominant frequencies in each signal as a function of time.

There exist several time-frequency distributions that have been researched in the context of many interesting applications. The most notable ones include:

- The spectrogram and its variations, such as the mel-spectrogram.
- The mel-frequency cepstral coefficients.
- The Wigner-Ville distribution.

The Wigner-Ville distribution essentially computes the Fourier transform of the signal's Ambiguity Function (AF), where:

$$AF(\tau; t) = x(t+\tau/2)x^*(t-\tau/2) \qquad (1)$$

The AF is a general representation of the signal's autocorrelation function. An advantage of the Wigner-Ville distribution is that is does not suffer from leakage effects, in contrast to the spectrogram. It is a quadratic transform and not a linear one, so that it creates cross terms when the signal is comprised of multiple dominant frequency components. The issue of cross terms is solved with an extended form of the Wigner-Ville distribution, the pseudo Wigner-Ville distribution. The latter is the one we used in this research. [13]

The Wigner-Ville distribution was chosen for three reasons. First, it provides more details in the time-frequency domain than the mainstream techniques like the spectrogram. Second, it was interesting to investigate its potential use for sound classification, since our literature research showed that most approaches neglected the Wigner-Ville distribution in favor for the spectrogram and MFCCs. Third, there exist papers that showcase the potential of the distribution for many interesting use cases. Some interesting examples include arrythmia detection [14], classification of bird-song syllables [15], image segmentation [16], and seismic applications [17].

## 5. Technical Approach

Based on the motivations and ideas described above, we formulated a processing plan that is comprised of the following steps:

1. Averaging the sound signals in case more than one sources of sound are available.
2. Filtering and subsequently down-sampling the signals using frequency filters. This step helps in focusing the analysis on the most relevant frequency range and reduces the computational time and power needed.
3. Transforming the signal to its analytical format, using the Hilbert transform. This step helped in removing the aliasing of frequencies that would be present without performing it.
4. Transforming the signals to one-channel images in the time-frequency domain using the pseudo Wigner-Ville distribution.
5. Classification of the Wigner-Ville arrays with the use of a convolutional neural network of our own design, that makes use of single-channel convolutions, motivated by the classification of grayscale images.

## 6. Results

The results in this section are divided according to what dataset was used for training and testing the classification model, as well as the associated use case.

### 6.1 UrbanSound8K

We built a classifier for the sounds present in the UrbanSound8K dataset. As mentioned in the previous chapter, the UrbanSound8K is a dataset open-sourced by the New York University, which contains around 8700 audio clips from 10 different categories of urban sounds. Most of the sound clips have a duration of about 4 seconds. Interestingly, the researchers mention in their paper [7] that, according to their findings, a duration of 4 seconds is enough to classify sound excerpts. In Figure 1, the number of audio clips available per class is depicted. From the figure, we see that most classes comprise more than 1000 sound samples, with siren, car horn, and gunshot being the only ones with a number lower than 1000, at a minimum of around 390 sound clips for gun shot.

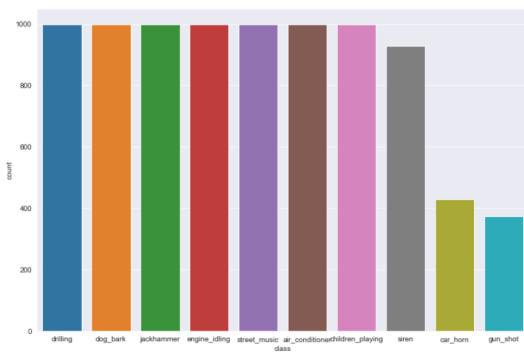

**Figure 1 - Number of sound clips per class in the UrbanSound8K dataset**

For this dataset we used the preprocessing steps as described in the Technical Approach section. Regarding the specifics of the implementation, for the down-sampling, following an approach of trial and error, we chose a sampling rate of 4 kHz, essentially focusing on the frequencies up to 2kHz. Using the technique of signal down-sampling, we also reduced the computation time of the Wigner-Ville distribution. Finally, after we transformed the sound signal to the Wigner-Ville distribution representation, we downsized the 2D image arrays to a dimension of 300x300 pixels using interpolation. The reason for that was that the original dimension of the arrays proved to be computationally intensive to serve as input for the Neural Network.

A successful architecture for the Neural Network is depicted in Figure 2. As described in the Technical Appoach section, the architecture is motivated by the VGG architecture for image classification; hence, it makes use of Convolutional and MaxPooling layers for feature learning and a fully-connected part for the final classification, as well. Moreover, incremental experimentation resulted in optimal hyperparameters for numbers of layers, filter kernels, depth of layers, and weights of dropout layers that are used to reduce overfitting. This experimentation started with an architecture that is documented to be successful in literature and gradually changed according to the results we observed during our experiments.

```
Net(
  (conv1): Conv2d(1, 16, kernel_size=(3, 3), stride=(1, 1), padding=(1, 1))
  (conv2): Conv2d(16, 32, kernel_size=(3, 3), stride=(1, 1), padding=(1, 1))
  (conv3): Conv2d(32, 64, kernel_size=(3, 3), stride=(1, 1), padding=(1, 1))
  (pool): MaxPool2d(kernel_size=2, stride=2, padding=0, dilation=1, ceil_mode=False)
  (fc1): Linear(in_features=87616, out_features=500, bias=True)
  (fc2): Linear(in_features=500, out_features=10, bias=True)
  (dropout): Dropout(p=0.25, inplace=False)
)
```

**Figure 2 - A neural network architecture used for the classification of the UrbanSound8K dataset**

Examples of the arrays of each class that served as input for the training of the neural network are shown in Figure 3. The distinctive patterns of the different classes of sounds are obvious to the naked eye. A promising initial signal that transforming to Wigner Ville distribution and performing classification on basis of that can be a successful strategy.

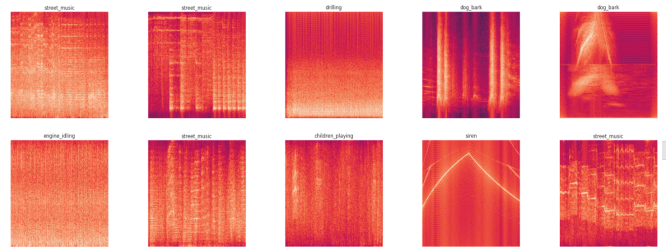

**Figure 3 - Preprocessed Wigner-Ville arrays that serve as input to the neural**

network architecture – the different patterns of different classes of sound are evident with a naked eye – the most prominent frequencies for each audio clip can be seen in yellow

The data was split in an 80/20 ratio, thus 80% of the data for training and 20% of the data for testing. The results of the validation on the test set are shown in Figure 4. Most of the sound excerpts were classified correctly by the Neural Network. It is interesting to note that classes such as drilling and jackhammer showed misclassifications, which makes sense, since their time-frequency distributions are very similar.

In Figure 5, the accuracy metrics for all the classes are shown. Overall, the neural network achieves an accuracy of 76%, a successful result in a 10-class random-guessing baseline of 10%.

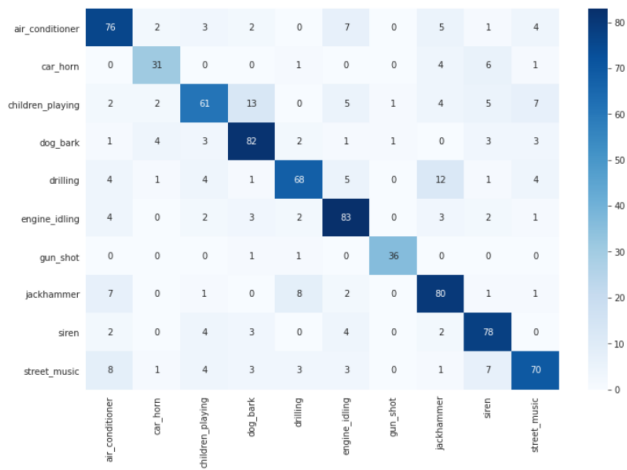

Figure 4 - Confusion matrix for the test set of an 80/20 split of the data – the majority of classes is correctly classified by the neural netwo

```
             precision    recall  f1-score   support

          0       0.73      0.76      0.75       100
          1       0.76      0.72      0.74        43
          2       0.74      0.61      0.67       100
          3       0.76      0.82      0.79       100
          4       0.80      0.68      0.74       100
          5       0.75      0.83      0.79       100
          6       0.95      0.95      0.95        38
          7       0.72      0.80      0.76       100
          8       0.75      0.84      0.79        93
          9       0.77      0.70      0.73       100

   accuracy                           0.76       874
  macro avg       0.77      0.77      0.77       874
weighted avg      0.76      0.76      0.76       874
```

Figure 5 - Accuracy metrics for all the classes of the UrbanSound8K dataset – an overall accuracy of 76% is achieved, a very successful result in a 10-class random-guessing baseline of 10%

Finally, we experimented with using the trained model to perform inference on an artificially-created streaming sound. The sound begins with children playing, continues to a dog barking, and finally returns to children playing. The signal is divided into 4-second overlapping windows with a stride of 1 second (first row of images in Figure 6), then transformed individually to the Wigner-Ville distribution (second row of images in Figure 6), which serve as input to the trained neural network that performs the inference successfully, as depicted in the third row of images in Figure 6.

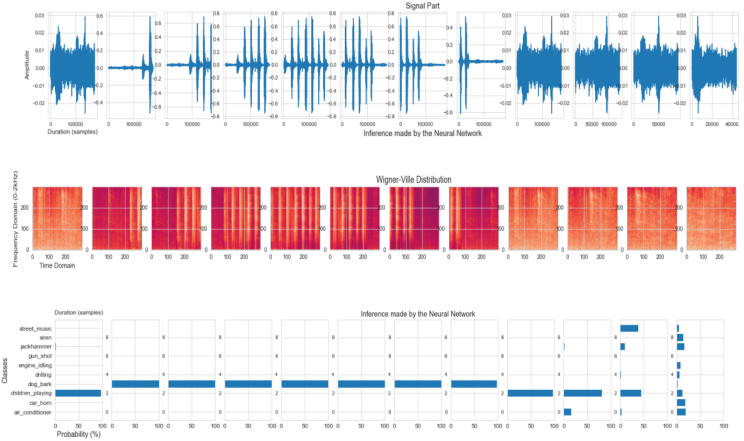

Figure 6 - Example of inference on streaming sound - the signal begins with children playing, passes to a dog barking, and finally returns to children playing. The signal is divided into 4-second overlapping windows with a stride of 1 second (first row of images), then transformed individually to the Wigner-Ville distribution (second row of images), which serve as input to the trained neural network that performs the inference successfully, as depicted in the third row of images.

## 6.2 Military Vehicle Classification

The goal of the classification of the two military vehicles in the military vehicle dataset was to create a proof of concept for classifying sounds of the engines coming from different types of vehicles. We split the data into an 80/20 train-test split, resulting in 311 entries in the training and 78 entries in the test set.

We followed the same steps explained before to preprocess the data for neural network training. In this dataset, we did not use down-sampling, since the data was already recorded at a low sampling rate of 4960Hz. Figure 7 shows examples of preprocessed signals that served as input for the training of the neural network. The most prominent frequencies for each signal are depicted in yellow. On top of each image, the type of vehicle that passed by the sound sensor is indicated.

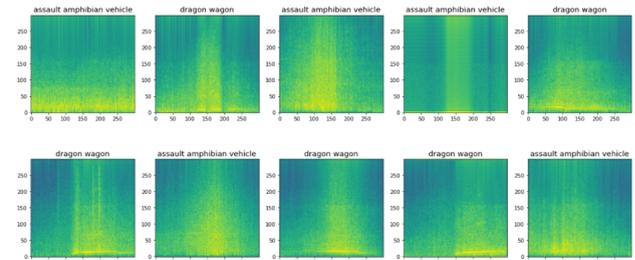

Figure 7 - Military Vehicle Classification dataset - examples of preprocessed signals that served as input for the training of the neural network – the most prominent frequencies for each signal are shown in yellow

Built with a similar architecture as the one shown in Figure 23, using the neural network we could classify the images, and thus the corresponding sounds, with an accuracy of 86% after 60 training epochs, a considerable improvement to a 50% baseline of random guessing between the two classes of vehicles.

*6.3 ESC-50*

We applied the same technical approach to classify the 50 classes of common sounds present in the ESC-50 dataset. The goal was to identify if there is potential in building a classifier with our approach on this broad dataset. Data was down-sampled at a sampling rate of 4 kHz, essentially focusing on the frequency range from 0 to 2kHz.

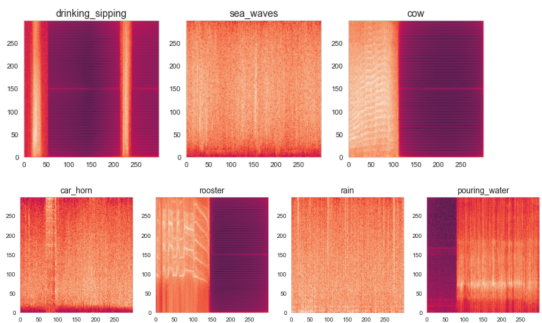

Figure 8 - Preprocessed signals from the ESC-50 dataset that served as input for training the neural network.

The dataset is already split in 5 folds for the purpose of cross-validation. Based on this, we trained models on 4 folds and tested them on the fifth fold that was left out from the training dataset at each point. This resulted in an average accuracy of 25% on the unseen folds, a significant improvement to random guessing 50 classes at 2% accuracy.

*6.4 Note on maximization of accuracy*

At this point, it is important to note that the optimization efforts made for the maximization of accuracy of the different models mentioned above was minimal. The focus was geared towards building proof of concepts fast that showcase the possible application of the approach in different interesting use cases. The accuracy has potential to increase if further efforts are made in the context of dataset creation/collection and hyperparameter tuning. This kind of optimization was not relevant at this phase of the research.

## 7. Conclusions

Our conclusions are summarized as follows:

- The use of sound sensors to detect events is possible with such high accuracy that it can be used for practical purposes.
- We gathered extensive documentation on open-source sound datasets and state-of-the-art research and technical approaches. It is important to note that the quality and number of open-source datasets in the context of urban sounds is limited, especially when compared to image and other structured datasets.
- Sound sensors are GDPR-compliant and less intrusive than sensors such as cameras.
- We designed a technical approach which transforms the sound signals to the Wigner-Ville time-frequency distribution. These transformed signals are treated as images and classified with the use of Convolutional Neural Networks.
- We validated positively the technical approach through several proof-of-concepts based on open-source datasets.

## 8. Future Work

There is undeniable potential in the classification of sound signals using AI techniques. What is missing currently and should drive future work is the application of such techniques on real-world datasets and applications, as well as the dissemination of such work in the form of open-source datasets and research papers, to further foster research in the field.

## 9. Acknowledgements

I would like to thank my supervisors Stef van Eijndhoven and Peter Duin for their continuous support and inspiration, as well as my colleagues at the Jheronimus Academy of Data Science (JADS) for the fruitful and engaging working environment.